\documentclass{pasj00}

\begin{document}
\SetRunningHead{D. Nogami et al.}{VW CrB in Outburst}

\Received{2002/00/00}
\Accepted{2002/00/00}

\title{Photometric Observations of an SU UMa-type Dwarf Nova VW~Coronae
Borealis during Outbursts}

\author{Daisaku \textsc{Nogami}}
\affil{Hida Observatory, Kyoto University,
       Kamitakara, Gifu 506-1314}
\email{nogami@kwasan.kyoto-u.ac.jp}
\author{Makoto \textsc{Uemura}, Ryoko \textsc{Ishioka}, Taichi \textsc{Kato}}
\affil{Department of Astronomy, Faculty of Science, Kyoto University,
Sakyo-ku, Kyoto 606-8502}
\author{Jochen \textsc{Pietz}}
\affil{Nollenweg 6, 65510 Idstein, Germany}


\KeyWords{accretion, accretion disks
          --- stars: novae, cataclysmic variables
          --- stars: dwarf novae
          --- stars: individual (VW CrB)}

\maketitle

\begin{abstract}

 We report the photometric observations of an SU UMa-type dwarf nova VW
 CrB during two superoutbursts in 2001 and 2003 and a normal outburst in
 2003.  Superhumps with a period of 0.07287(1) d were observed during
 the 2003 superoutburst.  The change rate of the superhump period was
 positive.  During the normal outburst, there are some hint of
 modulation up to a 0.2-mag amplitude.  However, any periodicity was not
 found.  The recurrence cycles of the normal outburst and the
 superoutburst, and the distance were estimated to be $\ge50$ d,
 $270\sim500$ d, and 690$^{+230}_{-170}$ pc, respectively.  These
 recurrence cycles are usual values for an SU UMa-type dwarf nova having
 this superhump period.  The superhump period of VW CrB was the longest
 among those of the SU UMa stars with positive derivatives of the
 superhump period.  The coverage of our observations was, however, not
 enough, and the variation of the $P_{\rm SH}$ change rate of VW CrB is
 still unknown.  A superhump regrowth and a brightening were seen near
 the end of the plateau phase.  Measuring the deviation of the start
 timings of the brightening and the superhump regrowth ($>$2 days in VW
 CrB) will be a key to reveal the mechanism of these phenomena.

\end{abstract}

\section{Introduction}

This variable star was discovered by \citet{ant96vwcrb}, and first
designated as Var21 CrB.  He found 9 outbursts from the Moscow
collection plates obtained between JD 2441805 and 2444042, and suggested
its dwarf nova nature showing two kinds (long and short) of outbursts.
Based on this report, \citet{NameList73} gave the permanent
variable-star name of VW CrB in the 73rd Name List.  \citet{nov97vwcrb}
caught superhumps with a period of 0.0743(6) d during a long outburst in
1997 May, establishing that VW CrB is an SU UMa-type dwarf nova.  A
spectrum taken by \citet{liu99CVspec1} during an outburst showed a blue
continuum having broad Balmer absorption lines with a central emission
core and He \textsc{ii} absorption, which is a typical spectrum for an
outbursting dwarf nova.  Na D absorption was also marginally detected.

VW CrB is identified with USNO B1.0 1231-0276740 ($B1=20.35, R1=19.55$).
The ROSAT sources, 1RXP J160003+3311.0 and 2RXP J160003.9+331115 are
X-ray counterparts.  No infrared counterpart was found in the 2
Micron All Sky Survey (2MASS) \citep{hoa02CV2MASS}.

\section{Observation}

We observed VW CrB during three outbursts in 2001 June-July, 2003 April,
and 2003 July-August at Kyoto University, Ouda Station, and Hida
observatory in Japan, and in Erfstadt in Germany.  The observation log
is summarized in table \ref{tab:log}.

\begin{table*}
\caption{Log of observations.}\label{tab:log}
\begin{center}
\begin{tabular}{ccrcccccc}
\hline\hline
\multicolumn{3}{c}{Date} & HJD-2400000 & Exposure & Frame  & filter & Instrument$^{\dagger}$ \\
         & &             & Start--End & Time (s) & Number &   & &  \\
\hline
2001 & July & 2 & 52093.105--52093.202 & 30 & 223 & no & A \\
     &      & 3 & 52094.082--52094.214 & 30 & 147 & no & B \\
     &      &   & 52094.109--52094.215 & 30 & 221 & no & A \\
\\
2003 & April & 9 & 52739.148--52739.261 & 10 & 550 & $R_{\rm c}$ & C \\
     &       &10 & 52740.151--52740.220 & 10 & 309 & $R_{\rm c}$ & C \\
     &       &12 & 52742.126--52742.196 & 30 &  56 & $R_{\rm c}$ & C \\
\\
2003 & July  &26 & 52847.024--52847.240 & 30  & 483 & no & A \\
     &       &28 & 52849.358--52849.499 & 60  & 125 & no & D \\
     &       &29 & 52850.350--52850.474 & 60  & 100 & no & D \\
     &       &31 & 52851.955--52852.137 & 20  & 490 & no & E \\
     &       &   & 52851.979--52852.109 & 30  & 306 & no & A \\
     &       &   & 52852.350--52852.475 & 60  & 109 & no & D \\
     & August& 1 & 52853.364--52853.425 & 60  &  22 & no & D \\
     &       & 2 & 52854.194--52854.205 & 30  &   7 & no & A \\
     &       &   & 52854.350--52854.533 & 60  & 145 & no & D \\
     &       & 3 & 52855.349--52855.483 & 60  &  63 & no & D \\
     &       & 4 & 52856.352--52856.465 & 60  &  46 & no & D \\
     &       & 5 & 52857.346--52857.471 & 60  &  69 & no & D \\
     &       & 6 & 52858.356--52858.477 & 60  &  81 & no & D \\
     &       & 7 & 52859.355--52859.476 & 60  &  64 & no & D \\
     &       & 9 & 52861.338--52861.492 & 60  &  95 & no & D \\
     &       &10 & 52862.354--52862.469 & 60  &  69 & no & D \\
     &       &11 & 52863.345--52863.471 & 60  &  83 & no & D \\
     &       &12 & 52864.347--52864.470 & 60  &  75 & no & D \\
     &       &14 & 52866.373--52866.380 & 60  &   2 & no & D \\
\hline
\multicolumn{8}{l}{A: 30cm Tel. + SBIG ST-7E (Kyoto), B: 25cm Tel. +
 SBIG ST-7E (Kyoto),} \\
\multicolumn{8}{l}{C: 60cm Tel. + PixelVision (SITe SI004AB) (Ouda),
 D: 20cm Tel. + SBIG} \\
\multicolumn{8}{l}{ST-6B (Erfstadt), E: 60cm Tel. + PixCellent S/T
 00-3194 (SITe 003AB) (Hida)} \\
 \end{tabular}
\end{center}
\end{table*}

The Kyoto frames and Ouda frames were processed by the PSF photometry
package developed by one of the authors (TK) after dark-subtraction and
flat-fielding.  All the frames obtained at Hida were reduced by the
aperture photometry package in IRAF\footnote{IRAF is distributed by the
National Optical Astronomy Observatories for Research in Astronomy,
Inc. under cooperative agreement with the National Science Foundation.},
after de-biasing and flat-fielding.  The Erfstadt frames were
dark-subtracted, flat-fielded, and analyzed with a PSF photometry
package.

\begin{figure}
  \begin{center}
    \FigureFile(84mm,84mm){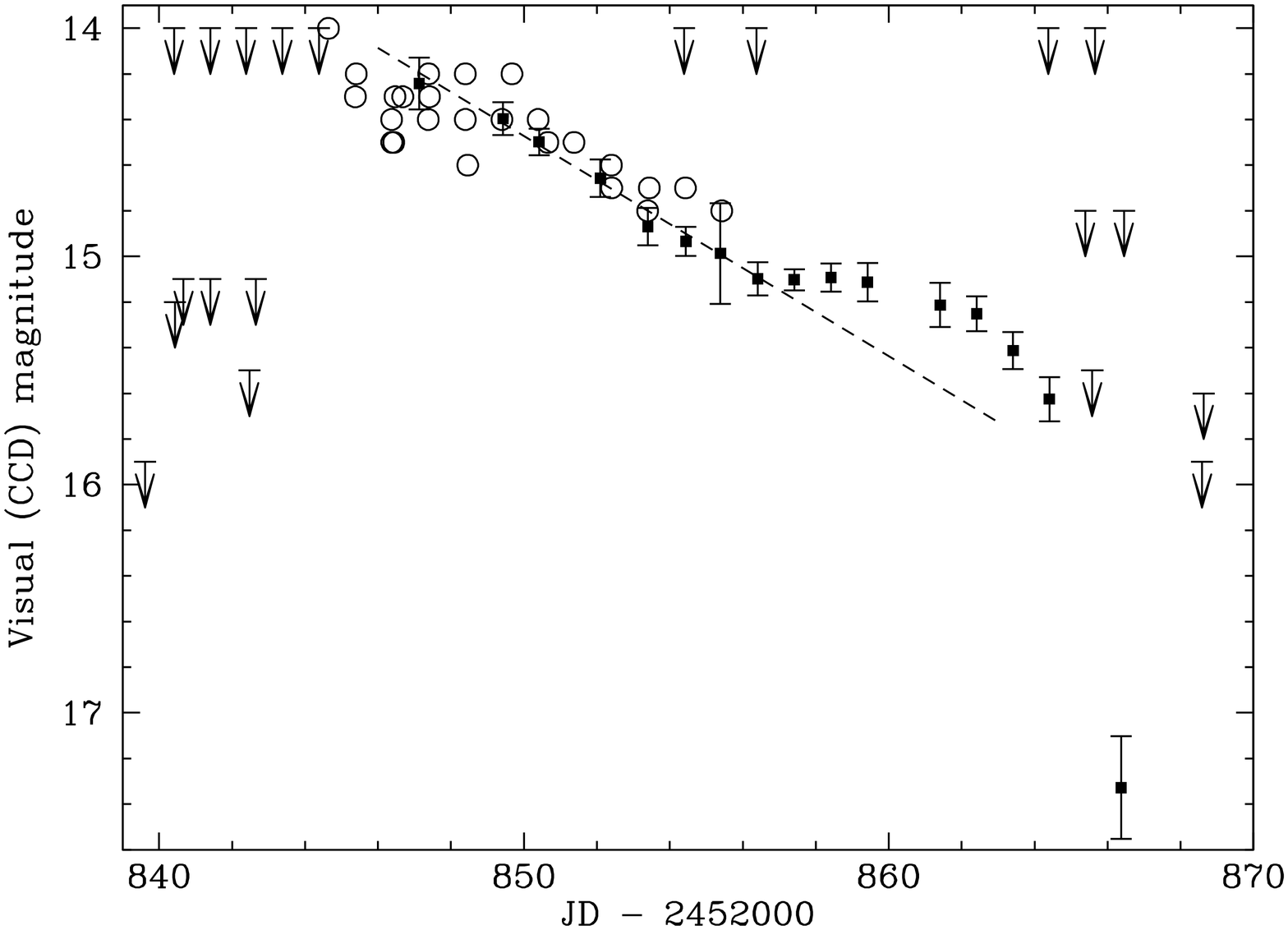}
  \end{center}
 \caption{The long-term light curve of the 2003 July-August outburst.
 The abscissa is HJD - 2452000, and the ordinate is Visual (CCD)
 magnitude.  The open circles and the lower arrows are the visual
 observations and the upper limits reported to VSNET.  The filled
 squares are daily average magnitudes of our data.  This outburst lasted
 about 20 days.  The dashed line represents the exponential decay trend
 of 0.097(4) mag d$^{-1}$ which was obtained using the daily averaged
 magnitudes before HJD 2452857.  It should be noted that VW CrB rose
 from the exponential decay after 14th day of this outburst.
 } \label{fig:0307long}
\end{figure}

The magnitudes of VW CrB in the Erfstadt frames were measured
differentially from the local comparison star USNO B1.0 1232-0272156,
which is the star of ID 2 ($V=14.044(2)$; $B-V=0.634(3)$) in the Arne
Henden's
sequence\footnote{$\langle$ftp://ftp.nofs.navy.mil/pub/outgoing/aah/sequence/sumner/ \\
vwcrb.seq$\rangle$}.
When the long-term light curve of the 2003 July outburst was drawn
(figure \ref{fig:0307long}), we shifted the zero point of these
differential magnitudes to fit them with the visual estimations reported
to VSNET \citep{VSNET}.  The differential magnitudes in the other frames
were adjusted to be in accordance with the Erfstadt data.

The 1-sigma error for each differential magnitude varied from 0.01 mag
to over 0.1 mag, depending on the target brightness and the sky
condition.

\section{Result}

In this section, we describe the observational results in detail in the
order of the 2003 July-August outburst, the 2001 June-July outburst, and
the 2003 April outburst, since the 2003 July-August outburst was best
covered and, therefore, most informative.

\subsection{The 2003 July-August Outburst}

The whole light curve of the 2003 July-August outburst is exhibited in
figure \ref{fig:0307long}.  This outburst was found by Mike Simonsen at
JD 2452844.639.  Our observations were started about 2.4 days later than
his report to VSNET, and within 4 days after onset of the outburst.  The
outburst duration was 20 days.  The dashed line represents the trend of
exponential decline of 0.097(4) mag d$^{-1}$ which was calculated by
linear regression of our data between JD 2452847 and 2452856.  From JD
2452857, the decline suddenly stopped, which is a `brightening' just
before the following rapid decline.  This phenomenon has been first
systematically investigated by \citet{kat03hodel} recently.

\begin{figure}
  \begin{center}
    \FigureFile(84mm,84mm){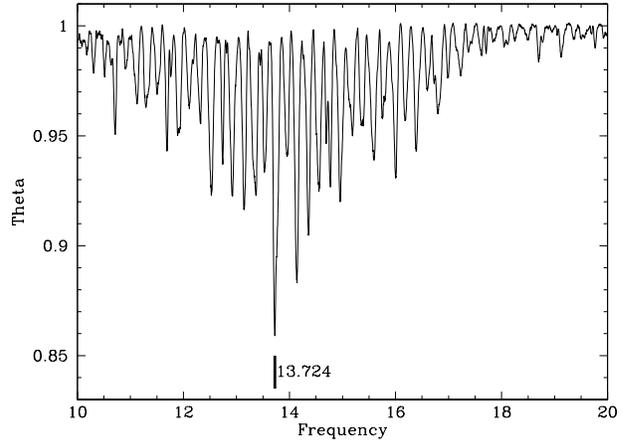}
  \end{center}
 \caption{Theta diagram of a PDM analysis to measure the superhump
 period from the data between JD 2452847 and 2452866.  The abscissa is
 the frequency in the unit of cycle d$^{-1}$.  The best estimated
 superhump period is 0.07287(1) d.
 } \label{fig:0307pdm}
\end{figure}

Our first run showed that superhumps had already fully developed by HJD
2452847.0 within 4 days from the outburst onset, which confirmed the SU
UMa nature of this star.  A period analysis by the Phase Dispersion
Minimization (PDM) method \citet{PDM} was performed for the data between
JD 2452847 and 2452866 after subtraction of the general decline trend
including the final `brightening'.  Figure \ref{fig:0307pdm} gives the
resultant theta diagram, which clearly shows the best estimated
superhump period ($P_{\rm SH}$) to be 0.07287(1) d.  The error was
calculated by the Lafler-Kinman methods \citep{fer89error}.

\begin{figure}
  \begin{center}
    \FigureFile(84mm,84mm){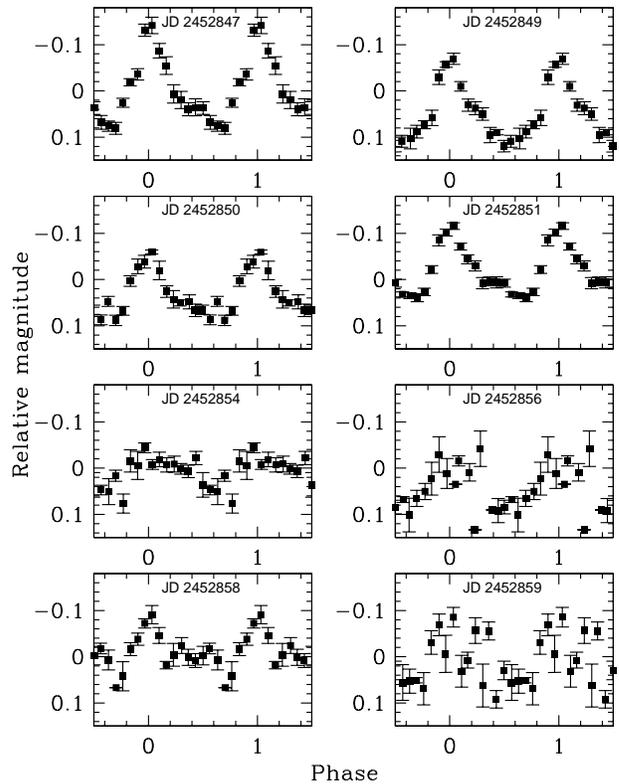}
  \end{center}
 \caption{Daily averaged superhump light curves folded by the superhump
 period of 0.07287 d.  The superhumps had fully developed to the
 amplitude of 0.24 mag within 4 days from the onset of the
 superoutburst.  The amplitude gradually became smaller to 0.10 mag
 till JD 2452854, then again developed to 0.14 mag on JD 2452858.  The
 superhump profile was smeared by the large observation error owing to
 thin clouds on JD 2452859.
 } \label{fig:0307shshape}
\end{figure}

The daily evolution of the superhump is depicted in figure
\ref{fig:0307shshape}.  The superhump amplitude was 0.24 mag on JD
2452847, and gradually decayed to 0.10 mag on JD 2452854.  After that,
the amplitude again grown to 0.14 mag by JD 2452858, though the large
error due to thin clouds smeared the superhump profile on JD 2452859.  A
clear second peak is seen around the phase = 0.4 on JD 2452858.

\begin{figure}
  \begin{center}
    \FigureFile(84mm,84mm){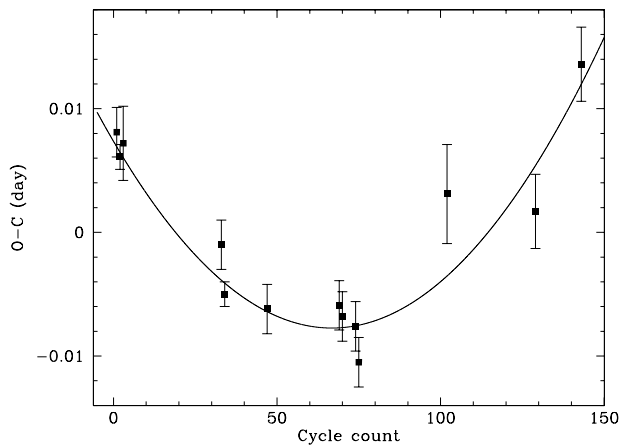}
  \end{center}
 \caption{$O-C$ diagram on the superhump maximum timings (see table
 \ref{tab:shmax} and equation (1)).  The solid line is
 obtained by a quadratic polynomial fitting to $O-C1$ (equation (2)),
 indicating a positive derivative of the superhump period.
 } \label{fig:0307shmax}
\end{figure}

The superhump maximum timings measured by eye were listed in table
\ref{tab:shmax}.  The cycle count ($E$) was set to be 1 at the first
superhump maximum.  Linear regression yields a following equation on the
maximum timings:
\begin{equation}
 C1 = 7.294 (\pm 0.005) \times 10^{-2} \cdot E - 0.019 (\pm 0.004).
\end{equation}
The difference of the observed timings from the calculated ones
systematically changed as shown in figure \ref{fig:0307shmax}.  A
quadratic polynomial fitting to $O-C1$ yields:
\begin{eqnarray}
 C2 & = & 3.39 (\pm 0.31) \times 10^{-6} \cdot (E-70)^2 \nonumber \\
    &   & + 2.2 (\pm 1.7) \times 10^{-5} \cdot (E-70)   \nonumber \\
    &   & - 0.0077 (\pm 0.0008).
\end{eqnarray}
The fittings were carried out using the non-linear least-squares
Marquardt-Levenberg algorithm.  The equation (2) indicates a positive
derivative of the superhump maximum timings, meaning that the superhump
period became longer.  The change rate of the superhump period is
$\dot{P}_{\rm SH}/P_{\rm SH} = 9.3 (\pm 0.9) \times 10^{-5}$.  Note
that these analyses are based on the data obtained by JD 2452857 and the
large error during the `brightening' phase unfortunately made the
behavior of the superhump period left unclear.

\begin{table}
\caption{Timings of the superhump maxima during the 2003 July-August
 superoutburst.}\label{tab:shmax}
\begin{center}
\begin{tabular}{rrr}
\hline\hline
HJD$-$2452847 &  E & O$-$C1$^*$ \\
\hline
  0.062(2) &   1 &  0.0081 \\
  0.133(1) &   2 &  0.0061 \\
  0.207(3) &   3 &  0.0072 \\
  2.387(2) &  33 & $-$0.0010 \\
  2.456(1) &  34 & $-$0.0050 \\
  3.403(2) &  47 & $-$0.0062 \\
  5.008(2) &  69 & $-$0.0059 \\
  5.080(2) &  70 & $-$0.0068 \\
  5.371(2) &  74 & $-$0.0076 \\
  5.441(2) &  75 & $-$0.0105 \\
  7.424(4) & 102 &  0.0031 \\
  9.392(3) & 129 &  0.0017 \\
 10.425(3) & 143 &  0.0136 \\
\hline
\multicolumn{3}{l}{$^*$ Using equation (1).  The unit} \\
\multicolumn{3}{l}{is day.}
\end{tabular}
\end{center}
\end{table}

\subsection{The 2001 June-July Outburst}

\begin{figure}
  \begin{center}
    \FigureFile(84mm,84mm){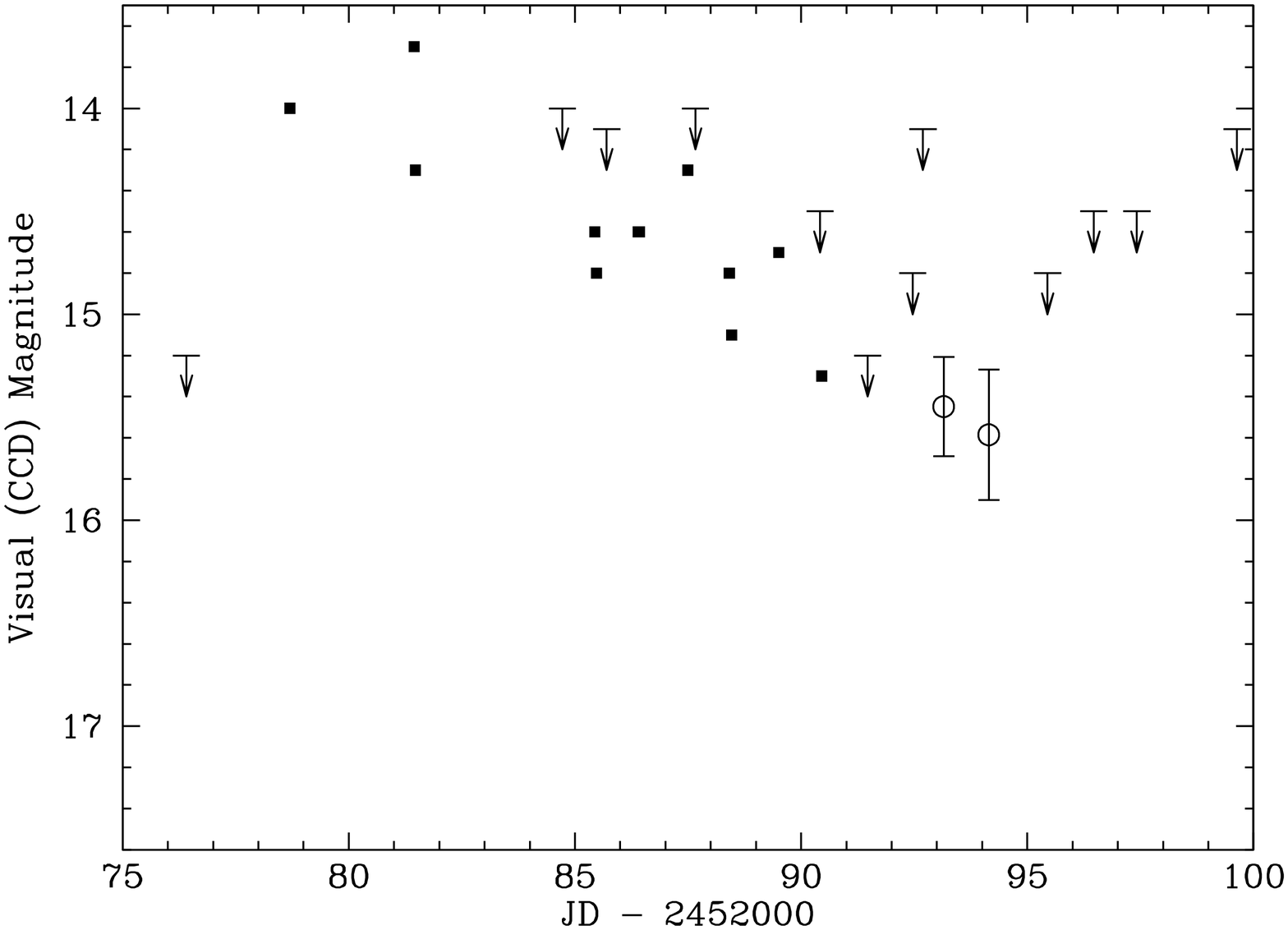}
  \end{center}
 \caption{Whole light curve of the 2001 June-July outburst.  This
 superoutburst lasted at least 15 days.  The zero point of our
 observations (open circles) have a large uncertainty of $\sim$0.3 mag.
 It is not clear whether the brightening just before the rapid decline
 occurred or not.
 } \label{fig:0107long}
\end{figure}

The long-term light curve of the 2001 July outburst is drawn in figure
\ref{fig:0107long}.  The long duration of $>$15 d proves that this
outburst was a superoutburst.  Since the uncertainty of the zero points
of each run is very large, $\sim$0.3 mag, it is difficult to judge
whether the final `brightening' occurred also in this superoutburst.

\begin{figure}
  \begin{center}
    \FigureFile(84mm,84mm){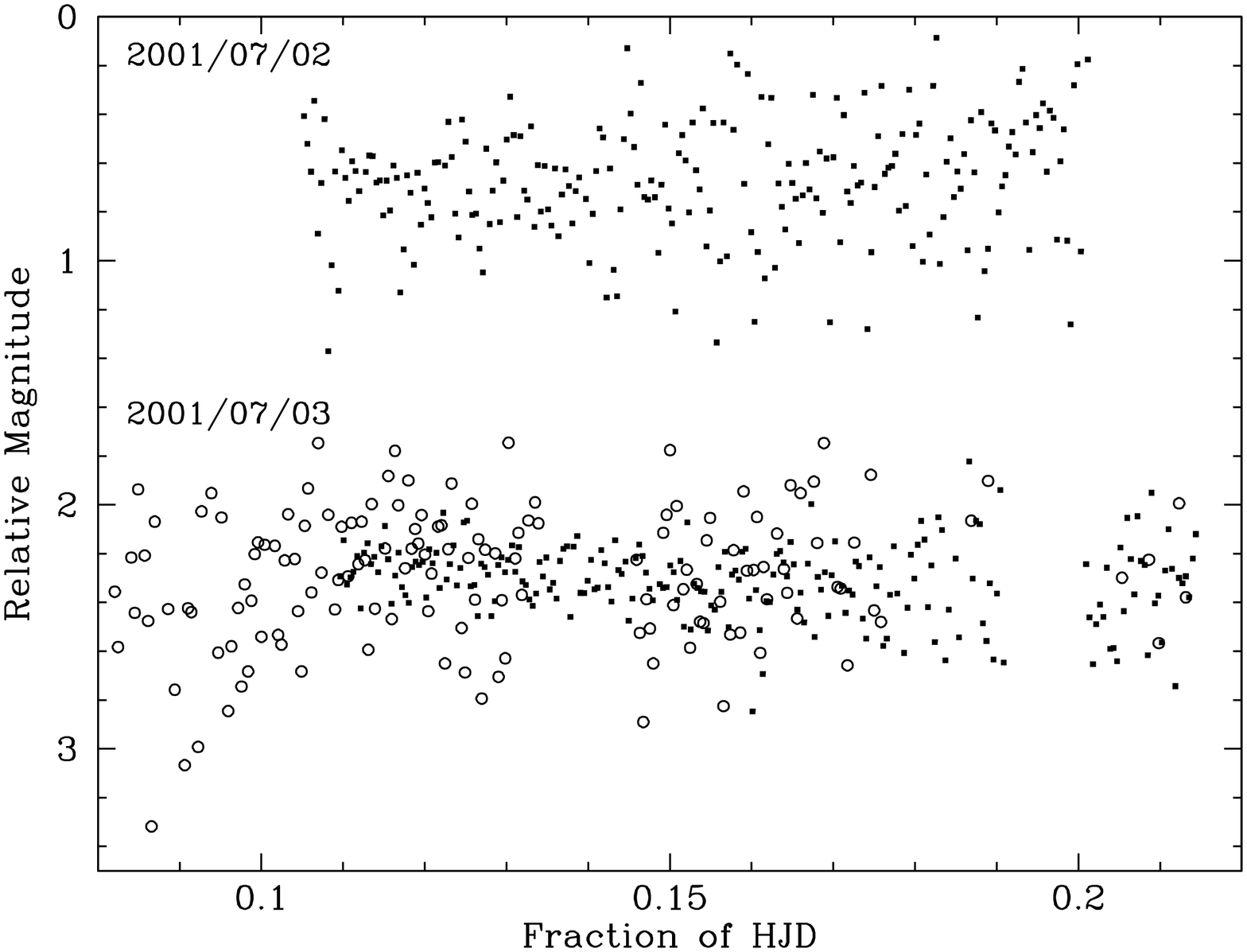}
  \end{center}
 \caption{Short-term light curve during the 2001 June-July outburst.
 The magnitudes are arbitrarily shifted each day.  The filled squares
 and open circles represent the data obtained by a 30cm telescope and
 25cm telescope, respectively.  The large errors and the short coverages
 prevent us from judging existence of superhumps with an amplitude
 smaller than 0.13 mag.
 } \label{fig:0107short}
\end{figure}

All the data points are plotted in figure \ref{fig:0107short}.  We could
not detect any periodic modulations including superhumps, which is
probably due to the large errors and the short coverages.

\subsection{The 2003 April Outburst}

\begin{figure}
  \begin{center}
    \FigureFile(84mm,84mm){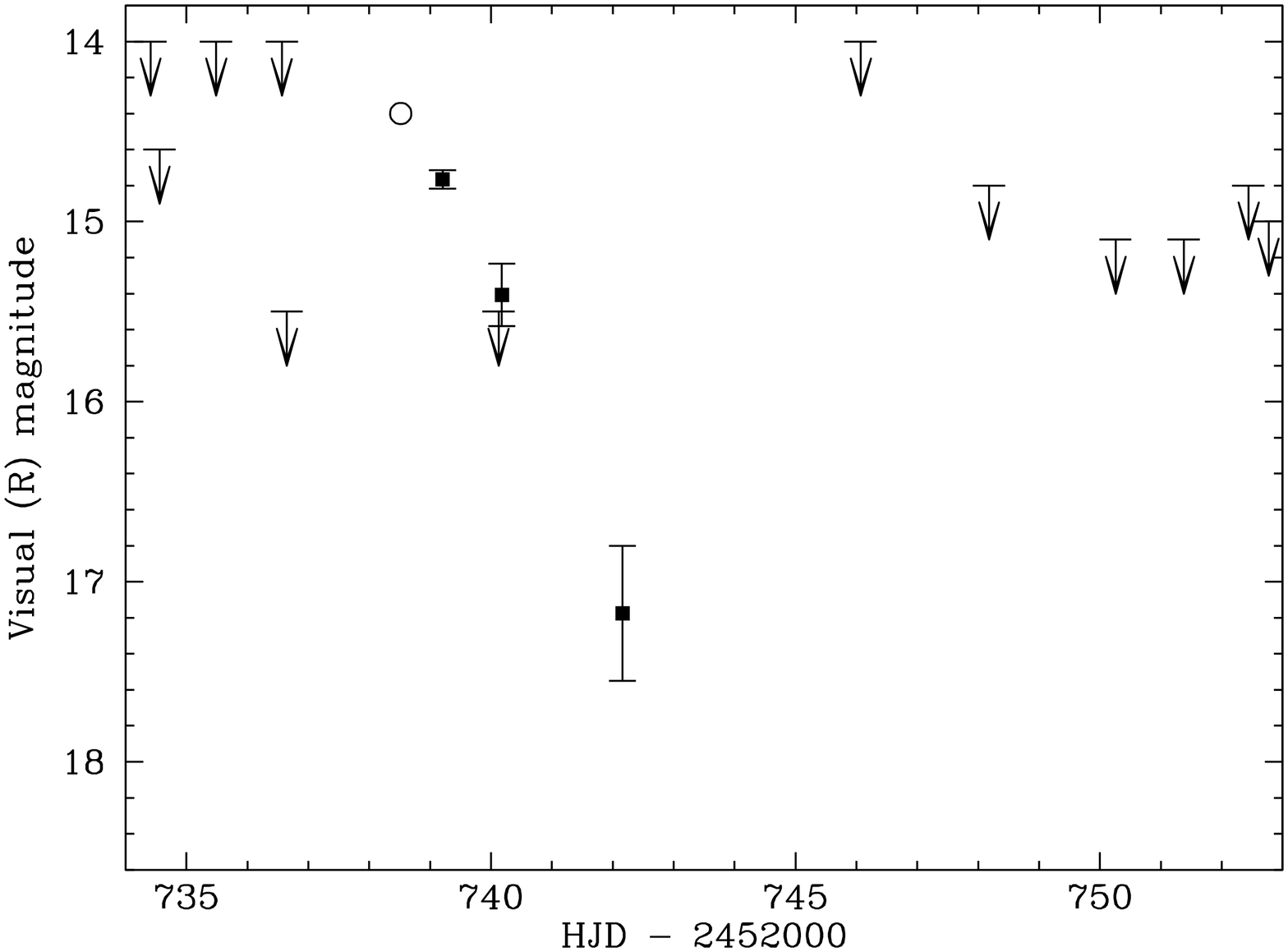}
  \end{center}
 \caption{Whole light curve of the 2003 April outburst.  The outburst
 duration was shorter than 5 days, suggesting a normal outburst.
 } \label{fig:0304long}
\end{figure}

Figure \ref{fig:0304long} displays the long-term light curve of the 2003
April outburst.  Since the duration of this outburst was shorter than 5
days, it was clearly a normal outburst.

\begin{figure}
  \begin{center}
    \FigureFile(84mm,84mm){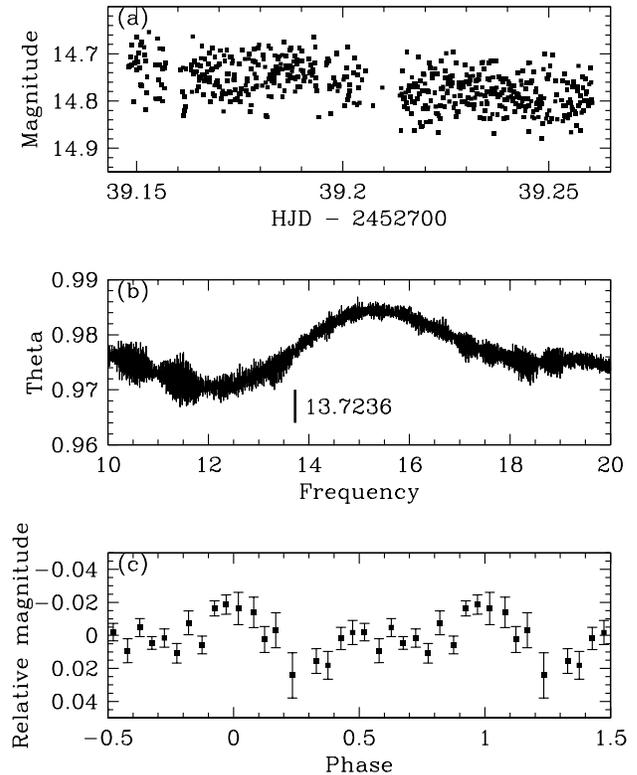}
  \end{center}
 \caption{(a) Light curve taken on JD 2452739.  (b) Theta diagram for
 the data shown in the upper panel after subtraction of the linear
 decline trend.  Any strong periodicity was not suggested.  The
 frequency of 13.7236 cycle d$^{-1}$ is the superhump frequency.  (c)
 Light curve folded by the superhump period after subtraction of the
 linear decline trend.  Although a plausible modulation emerges, its
 reality is fairly weak.
 } \label{fig:03040409}
\end{figure}

The light curve on JD 2452739 is shown in figure \ref{fig:03040409}a.
There are some hints of modulations.  After subtracting the declining
trend of 0.7 mag d$^{-1}$, we performed a PDM period analysis on the
data.  The resultant $\Theta$ diagram is figure \ref{fig:03040409}b,
where no strong signal of periodicity is seen.  If we make a light curve
folded by the superhump period after subtraction of the general decline
trend, a plausible light curve emerges (figure \ref{fig:03040409}c).
This periodicity is, however, most doubtful, since the $\Theta$ diagram
in figure \ref{fig:03040409}b does not point to that frequency, the full
amplitude of 0.04 mag is nearly equal to the 1-sigma error for each
point, and the coverage of the data is not enough for the period.

\section{Discussion}

\subsection{Outburst Behavior}

The outbursts reported to VSNET are summarized in table
\ref{tab:outburst}.

VW CrB has been relatively well monitored.  The maximum magnitude of
the superoutburst is rather bright, reaching $m_{\rm vis} = 14.0$, and
the superoutburst has long durations up to $>$23 d for an SU UMa-type
dwarf nova.  For these reasons, all the superoutbursts has been likely
caught after the start of monitoring by amateurs, except for during the
season when this variable is not observable.  Although the recurrence
cycle of the superoutburst (supercycle, $T_{\rm s}$) is not stable, it
is estimated to be $T_{\rm s} = 270\sim500$ d, assuming that one
superoutburst was missed between the 1998 September superoutburst and
the 2000 September one.

The normal outburst is $m_{\rm vis} = 14.4$ at the maximum, fainter than
in the superoutburst, and the duration is shorter than 5 d.  Then,
some normal outbursts have been missed.  Taking also into account the
outbursts observed by \citet{ant96vwcrb}, it is proper to estimate the
cycle of the normal outburst ($T_{\rm n}$) to be 50 d or more.

This supercycle and the ratio of the normal-outburst cycle to the
supercycle are ordinary values for a dwarf nova with $P_{\rm SH}=0.072$
d (see \cite{war95suuma,nog97sxlmi,kat03hodel}).

The superoutburst duration, up to $>$23 d as mentioned before, is
$\ge$1.5 times longer than those of usual SU UMa stars, and is
comparable of those of ER UMa \citep{kat95eruma,rob95eruma} and the
durations of the main outburst (until the dip) of superoutbursts in WZ
Sge stars (e.g. the 2001 superoutburst of WZ Sge, see
\cite{ish02wzsgeletter,pat02wzsge}).  In the current scheme of the
standard thermal-tidal disk instability model (for a review, see
\cite{osa96review}), the outburst enter the rapid decay phase when the
surface density becomes smaller than the critical value at the outer
edge of the contracting accretion disk and the cooling wave starts.
There are three possible conditions to make the outburst duration longer
in the disk instability model: 1) large mass is supplied during the
outburst (as the explanation for the ER UMa case, \cite{osa95eruma}), 2)
larger mass is stored in the disk by the outburst ignition (as the
explanation for the WZ Sge case, \cite{osa95wzsge}), and 3) the outburst
lasts until the disk become smaller.

\begin{table}
\caption{Previous outbursts.}\label{tab:outburst}
\begin{center}
\begin{tabular}{llrclrc}
\hline\hline
\multicolumn{3}{c}{Date$^*$} & JD & $V_{\rm max}$ & D$^\dagger$ & Type$^\ddagger$ \\
\hline
1996 & Jun. & 26 & 50261 & 15.4  &       & N  \\
1996 & Aug. & 13 & 50309 & 14.3: & $>$6  & S  \\
1997 & Feb. & 27 & 50506 & 15.0: & 4?    & N  \\
1997 & May  & 11 & 50579 & 14.1  & $>$20 & S  \\
1997 & Aug. & 20 & 50680 & 14.7  & 2     & N \\
1998 & Sep. & 27 & 51083 & 13.8  & $>$23 & S \\
1999 & Apr. & 25 & 51293 & 14.4  & 1?    & N \\
2000 & Apr. & 08 & 51643 & 14.6  & $>$2  & N \\ 
2000 & Sep. & 13 & 51800 & 13.9: & $>$11 & S \\
2001 & Jun. & 18 & 52078 & 13.7  & $>$15 & S \\
2002 & Jun. & 23 & 52449 & 14.2  & $>$18 & S \\
2003 & Apr. & 09 & 52738 & 14.4  & $<$5  & N \\
2003 & Jul. & 24 & 52844 & 14.0  & 20    & S \\
\hline
\multicolumn{7}{l}{$^*$ The discovery date.}\\
\multicolumn{7}{l}{$^\dagger$ Duration of the outburst in a unit of day.}\\
\multicolumn{7}{l}{$^\ddagger$ N: normal outburst, S: superoutburst.}
\end{tabular}
\end{center}
\end{table}

VW CrB, however, seems to have a normal mass transfer rate, judging from
its ordinary $T_{\rm s}$, and our data provides no hint of enhancement
of the mass transfer rate during the superoutburst.  The possibility 1)
is hence rejected.

Concerning the possibility 2), the amount of the mass stored by the
outburst ignition is related to physical parameters of the viscosity
in quiescence, the disk radius, the primary mass
(cf. \cite{osa95wzsge}).  These parameters, however, also have an effect
on $T_{\rm n}$ and $T_{\rm s}$.  Since the cycle lengths are normal
values for an SU UMa star, it is difficult to suppose unusual parameters
on these parameters.

Therefore, the possibility 3) is most plausible.  This possibility means
that the tidal torque strongly works on the disk gas.  Since the mass
ratio dominantly determines the tidal effect, it is quite important to
accurately measure the mass ratio of this system.

\subsection{Distance}

From the superhump period and the maximum magnitude of the normal
outburst, we can estimate the distance to VW CrB in the same method as
that used by \citet{kat02v592her}, \citet{kat03hodel} and
\citet{nog03qwser}.

\citet{war87CVabsmag} proposed a relationship between the absolute
magnitude at the outburst maximum and the orbital period $P_{\rm orb}$.
The orbital period of VW CrB has not yet been measured, we substitute
$P_{\rm SH}$ for $P_{\rm orb}$.  There is a well-known, tight
relationship between $P_{\rm orb}$ and $P_{\rm SH}$, and the error
introduced by this substitution is negligible, compared with those due
to other factors.  Absence of the eclipse indicates that the inclination
of this system is not so high, and the inclination effect to the
observed flux \citep{war86NLabsmag} should be very small.  The absolute
maximum magnitude is thus estimated to be $M_{V} = 5.20 (\pm 0.23)$,
using the equation proposed by \citet{war87CVabsmag}.  This maximum
magnitude should be compared by that of the normal outburst
\citep{kat02v592her}.  Then, assuming the apparent maximum magnitude
$m_{V} = 14.4 (\pm 0.4)$ from the data in table \ref{tab:outburst}, the
distance is guessed to be 690$^{+230}_{-170}$ pc.

The proper motion of the star identified with the VW CrB is listed in
the USNO B1.0 catalog as ($\mu_{\rm R.A.}, \mu_{\rm Dec}$) = ($-$6(10),
$-$26(7)) in a unit of mas yr$^{-1}$.  This distance is smaller than the
secure upper limit estimated using this proper motion and the maximum
expected velocity dispersion of CVs (\cite{har00DNdistance}).

The X-ray luminosity in the range of 0.5-2.5 keV of VW CrB is calculated
to be $\log L_{\rm X} = 31.0 \pm 0.2$, using the ROSAT data of the X-ray
counterparts and the distance in the same way by \citet{ver97ROSAT}.
This luminosity is a little higher than, but not far from, the average
value of SU UMa stars (see figures 6 and 8 in \cite{ver97ROSAT}).

\subsection{Superhump Period Change}

The superhump period had been believed to monotonically decrease, or
stay constant at least (e.g. \cite{war85suuma,pat93vyaqr}), before the
discovery of $P_{\rm SH}$ increase during the 1995 superoutburst in AL
Com \citep{nog97alcom}.  Following this, some SU UMa stars have been
confirmed to show similar behaviors: V485 Cen \citep{ole97v485cen},
EG Cnc \citep{kat03egcnc}, SW UMa \citep{sem97swuma, nog98swuma}, V1028
Cyg \citep{bab00v1028cyg}, WX Cet \citep{kat01wxcet}, HV Vir
\citep{kat01hvvir,ish03hvvir}, V592 Cas \citep{kat02v592cas}, WZ Sge
(\cite{pat02wzsge}; Ishioka et al. 2003, in preparation), EI Psc
\citep{uem02j2329}, V1141 Aql \citep{ole03v1141aql}, and KS UMa
\citep{ole03ksuma}.  Superhump-period decrease was explained by shrinkage
of the disk radius, and was regarded as a natural consequence of mass
depletion from the disk \citep{osa85SHexcess}.  The mechanism of the
$P_{\rm SH}$ increase phenomenon is, however, still an open question.

The superhump period of VW CrB is the longest among those of the SU
UMa-type dwarf novae with positive $\dot{P}_{\rm SH}$, listed above.
Noticing the concentration of these SU UMa stars near the orbital period
minimum, \citet{kat01hvvir} suggested that the period increase may be
related to a low mass ratio and/or a low mass transfer rate.  This may
be supported by MN Dra \citep{nog03var73dra}.  This star is considered
to have a high mass ratio and a high mass transfer rate, based on its
long $P_{\rm orb}$ of 0.105 d and short $T_{\rm s}$ of $\sim$60 d.  And
the $P_{\rm SH}$ derivative was $\dot{P}_{\rm SH}/P_{\rm SH} =
1.7\times10^{-3}$, about one order of magnitude larger than the largest
value ever know.

Recent observations have revealed that some SU UMa stars with a
relatively high mass ratio, a high mass transfer rate, and a long
orbital period have $\dot{P}_{\rm SH}/P_{\rm SH} \sim 0$ (e.g. BF Ara,
\cite{kat03bfara}).  Furthermore, \citet{ole03ksuma} most recently
proposed that SU UMa stars show a positive $\dot{P}_{\rm SH}/P_{\rm SH}$
around the midst of the plateau phase, and then change $\dot{P}_{\rm
SH}/P_{\rm SH}$ to be a minus value during a late phase of the
superoutburst.  Our present data at the late phase, unfortunately,
contain a large error, and can not be useful to make clear the evolution
of the superhump period throughout a superoutburst.

\subsection{Final `Brightening' and Superhump Regrowth}

VW CrB gave rise to a phenomenon of superhump regrowth and a final
brightening at the end of the plateau phase of the 2003 July-August
superoutburst, as described in section 3.1.  \citet{kat03hodel} did a
literature survey, and discussed the relationship between the final
brightening and the superhump regrowth.  The former were seen in many
short-$P_{\rm SH}$ systems (but few WZ Sge stars) and some long-$P_{\rm
SH}$ systems with long $T_{\rm s}$.  Dwarf novae having shown the latter
feature, except for V725 Aql, are more concentrated to the short end of
the $P_{\rm SH}$ distribution.  \citet{kat03hodel} gave a remark that
the appearance of superhump regrowth and brightening near the
termination of the superoutburst are phenomenologically coupled.

VW CrB is located at a slightly longer side out of the $P_{\rm SH}$
region of dwarf novae with such phenomena, and has a relatively short
$T_{\rm s}$ for a system with the brightening.  This dwarf nova may be
an important object to determine the critical value of the binary
parameters separating the SU UMa stars with/without the final
brightening and the superhump regrowth.

The superhump amplitude on JD 2452856 was already larger than on JD
2452854 (figure \ref{fig:0307shshape}), while the brightening seemed to
start on JD 2452856 (figure \ref{fig:0307long}).  To reveal the
connection and the mechanism of the final brightening and the superhump
regrowth, one of the keys will be more detailed observations in many SU
UMa stars which measure deviation of the start timings of the
brightening and the superhump regrowth.

\section{Summary}

The SU UMa nature of VW CrB was confirmed, and some parameters are
determined: the superhump period of $P_{\rm SH} =$0.07287(1) d, the
$P_{\rm SH}$ change rate of $\dot{P}_{\rm SH}/P_{\rm SH}
= 9.3 (\pm 0.9) \times 10^{-5}$, the supercycle of $T_{\rm s} =
270\sim500$ d, the cycle length of the normal outburst of $T_{\rm n}
\ge50$ d, the superoutburst duration up to $>$23 d, the distance of
690$^{+230}_{-170}$ pc, and the X-ray (0.5-2.5 keV) luminosity of
$\log L_{\rm X} = 31.0 \pm 0.2$.

The supercycle and the cycle length of the normal outburst are usual
values for an SU UMa star.  The superoutburst duration is very long, and
comparable to that of ER UMa and the duration of the main outburst of
superoutbursts in WZ Sge stars.  This may be due to a strong tidal
torque.  It is important to gauge the mass ratio.

The superhump period of VW CrB was the longest among those of the SU
UMa-type dwarf novae with positive derivatives of the superhump period.
Although positive-$\dot{P}_{\rm SH}$ systems have been supposed to have
short orbital periods, most recent observations show that the the sign
of $P_{\rm SH}$ change rate varies even during one superoutburst in some
SU UMa stars.  Since our observations at the late phase of the 2003
superoutburst was performed under bad conditions, the variation of the
$P_{\rm SH}$ change rate of VW CrB is a topic left for future
observations.

During the 2003 superoutburst, we observed regrowth of the superhump
amplitude and brightening near the end of the plateau phase.  These
features have been observed mainly in SU UMa stars having short
superhump periods.  The superhump period and the supercycle of VW CrB
are slightly longer and shorter, respectively, than those in such
systems.  Measuring deviation of the start timings of the
brightening and the superhump regrowth ($>$2 d in VW CrB) will help to
reveal the mechanism of the superhump regrowth and the final
brightening.

\vskip 3mm

The authors are thankful to amateur observers for continuous
reporting their valuable observations to VSNET.  Thanks are also to the
referee for the useful comments. This work is partly supported by a
Research Fellowship of the Japan Society for the Promotion of Science
for Young Scientists (MU and RI), and a grant-in-aid from the Japanese
Ministry of Education, Culture, Sports, Science and Technology
(No. 13640239, 15037205).


\begin{thebibliography}{}

\bibitem[Antipin(1996)]{ant96vwcrb}
  Antipin, S.~V.\ 1996, Inf. Bull. Variable Stars, 4343

\bibitem[Baba et~al.(2000)]{bab00v1028cyg}
  Baba, H., Kato, T., Nogami, D., Hirata, R., Matsumoto, K., \& Sadakane, K.\
  2000, \pasj, 52, 429

\bibitem[Fernie(1989)]{fer89error}
  Fernie, J.~D.\ 1989, \pasp, 101, 225

\bibitem[Harrison et~al.(2000)]{har00DNdistance}
  Harrison, T.~E., McNamara, B.~J., Szkody, P., \& Gilliland, R.~L.\ 2000, \aj,
  120, 2649

\bibitem[Hoard et~al.(2002)]{hoa02CV2MASS}
  Hoard, D.~W., Wachter, S., Clark, L.~L., \& Bowers, T.~P.\ 2002, \apj, 565,
  511

\bibitem[Ishioka et~al.(2003)]{ish03hvvir}
  Ishioka, R. {et~al.}\ 2003, \pasj, 55, 683

\bibitem[Ishioka et~al.(2002)]{ish02wzsgeletter}
  Ishioka, R. {et~al.}\ 2002, \aap, 381, L41

\bibitem[Kato et~al.(2003a)]{kat03bfara}
  Kato, T., Bolt, G., Nelson, P., Monard, B., Stubbings, R., Pearce, A.,
  Yamaoka, H., \& Richards, T.\ 2003a, \mnras, 341, 901

\bibitem[Kato, Kunjaya(1995)]{kat95eruma}
  Kato, T., \& Kunjaya, C.\ 1995, \pasj, 47, 163

\bibitem[Kato et~al.(2001a)]{kat01wxcet}
  Kato, T., Matsumoto, K., Nogami, D., Morikawa, K., \& Kiyota, S.\ 2001a,
  \pasj, 53, 893

\bibitem[Kato et~al.(2003b)]{kat03egcnc}
  Kato, T., Nogami, D., Matsumoto, K., \& Baba, H.\ 2003b, \pasj,
  in press (astro-ph/0310426)

\bibitem[Kato et~al.(2003c)]{kat03hodel}
  Kato, T., Nogami, D., Moilanen, M., \& Yamaoka, H.\ 2003c, \pasj,
  in press (astro-ph/0307064)

\bibitem[Kato et~al.(2001b)]{kat01hvvir}
  Kato, T., Sekine, Y., \& Hirata, R.\ 2001b, \pasj, 53, 1191

\bibitem[Kato, Starkey(2002)]{kat02v592cas}
  Kato, T., \& Starkey, D.~R.\ 2002, Inf. Bull. Variable Stars, 5358

\bibitem[Kato et~al.(2003d)]{VSNET}
  Kato, T., Uemura, M., Ishioka, R., Nogami, D., Kunjaya, C., Baba, H., \&
  Yamaoka, H.\ 2003d, \pasj, in press (astro-ph/0310209)

\bibitem[Kato et~al.(2002)]{kat02v592her}
  Kato, T., Uemura, M., Matsumoto, K., Kinnunen, T., Garradd, G., Masi, G., \&
  Yamaoka, H.\ 2002, \pasj, 54, 999

\bibitem[Kazarovets, Samus(1997)]{NameList73}
  Kazarovets, E.~V., \& Samus, N.~N.\ 1997, Inf. Bull. Variable Stars, 4471

\bibitem[Liu et~al.(1999)]{liu99CVspec1}
  Liu, Wu., Hu, J.~Y., Zhu, X.~H., \& Li, Z.~Y.\ 1999, \apjs, 122, 243

\bibitem[Nogami et~al.(1998)]{nog98swuma}
  Nogami, D., Baba, H., Kato, T., \& Nov\'{a}k, R.\ 1998, \pasj, 50, 297

\bibitem[Nogami et~al.(1997a)]{nog97alcom}
  Nogami, D., Kato, T., Baba, H., Matsumoto, K., Arimoto, J., Tanabe, K., \&
  Ishikawa, K.\ 1997a, \apj, 490, 840

\bibitem[Nogami et~al.(1997b)]{nog97sxlmi}
  Nogami, D., Masuda, S., \& Kato, T.\ 1997b, \pasp, 109, 1114

\bibitem[Nogami et~al.(2003a)]{nog03qwser}
  Nogami, D. {et~al.}\ 2003a, \pasj, in press (astro-ph/0310274)

\bibitem[Nogami et~al.(2003b)]{nog03var73dra}
  Nogami, D. {et~al.}\ 2003b, \aap, 404, 1067

\bibitem[Nov\'{a}k(1997)]{nov97vwcrb}
  Nov\'{a}k, R.\ 1997, Inf. Bull. Variable Stars, 4489

\bibitem[Olech(1997)]{ole97v485cen}
  Olech, A.\ 1997, Acta Astron., 47, 281

\bibitem[Olech(2003)]{ole03v1141aql}
  Olech, A.\ 2003, Acta Astron., 53, 85

\bibitem[Olech et~al.(2003)]{ole03ksuma}
  Olech, A., Schwarzenberg-Czerny, A., K\c{e}dzierski, P., Z{\l}oczewski, K.,
  Mularczyk, K., \& Wi\'{s}niewski, M.\ 2003, Acta Astron., 53, 175

\bibitem[Osaki(1985)]{osa85SHexcess}
  Osaki, Y.\ 1985, \aap, 144, 369

\bibitem[Osaki(1995a)]{osa95eruma}
  Osaki, Y.\ 1995a, \pasj, 47, L11

\bibitem[Osaki(1995b)]{osa95wzsge}
  Osaki, Y.\ 1995b, \pasj, 47, 47

\bibitem[Osaki(1996)]{osa96review}
  Osaki, Y.\ 1996, \pasp, 108, 39

\bibitem[Patterson et~al.(1993)]{pat93vyaqr}
  Patterson, J., Bond, H.~E., Grauer, A.~D., Shafter, A.~W., \& Mattei, J.~A.\
  1993, \pasp, 105, 69

\bibitem[Patterson et~al.(2002)]{pat02wzsge}
  Patterson, J. {et~al.}\ 2002, \pasp, 114, 721

\bibitem[Robertson et~al.(1995)]{rob95eruma}
  Robertson, J.~W., Honeycutt, R.~K., \& Turner, G.~W.\ 1995, \pasp, 107, 443

\bibitem[Semeniuk et~al.(1997)]{sem97swuma}
  Semeniuk, I., Olech, A., Kwast, T., \& Nalezyty, M.\ 1997, Acta
		       Astron., 47, 201

\bibitem[Stellingwerf(1978)]{PDM}
  Stellingwerf, R.~F.\ 1978, \apj, 224, 953

\bibitem[Uemura et~al.(2002)]{uem02j2329}
  Uemura, M. {et~al.}\ 2002, \pasj, 54, 599

\bibitem[Verbunt et~al.(1997)]{ver97ROSAT}
  Verbunt, F., Bunk, W.~H., Ritter, H., \& Pfeffermann, E.\ 1997, \aap, 327,
  602

\bibitem[Warner(1985)]{war85suuma}
  Warner, B.\ 1985, in Interacting Binaries, ed. P.~P. Eggelton \& J.~E.
  Pringle (Dordrecht: D. Reidel Publishing Company), ~367

\bibitem[Warner(1986)]{war86NLabsmag}
  Warner, B.\ 1986, \mnras, 222, 11

\bibitem[Warner(1987)]{war87CVabsmag}
  Warner, B.\ 1987, \mnras, 227, 23

\bibitem[Warner(1995)]{war95suuma}
  Warner, B.\ 1995, \apss, 226, 187

\end{thebibliography}
\end{document}